\documentclass{appolb}
\usepackage{epsfig}
\begin{document}

\title{$\beta$ DECAY STUDIES OF NEUTRON-RICH NUCLEI NEAR $^{52}$Ca
\thanks{Presented at the Zakopane Conference on Nuclear Physics,
September 1-7, 2008}}
\author{
H.L. Crawford$^{a,b}$, R.V.F. Janssens$^c$, P.F. Mantica$^{a,b}$, J.S. Berryman$^{a,b}$, R. Broda$^d$, M.P. Carpenter$^c$, B. Fornal$^d$, G.F. Grinyer$^b$, N. Hoteling$^{c,e}$, B. Kay$^c$, T. Lauritsen$^c$, K. Minamisono$^b$, I. Stefanescu$^{c,e}$, J.B. Stoker$^{a,b}$, W.B. Walters$^e$, S. Zhu$^c$
\address{
$^a$Dept. of Chemistry, Michigan State University, E. Lansing, MI 48824, USA\\
$^b$National Superconducting Cyclotron Laboratory, Michigan State University, E. Lansing, MI 48824, USA\\
$^c$Physics Division, Argonne National Laboratory, Argonne, IL 60439, USA\\
$^d$Institute of Nuclear Physics, Polish Academy of Sciences, PL-31342, Cracow, Poland\\
$^e$Dept. of Chemistry and Biochemistry, University of Maryland, College Park, MD 20742, USA}}
\maketitle

\begin{abstract}
The $\beta$-decay and isomeric properties of $^{54}$Sc, $^{50}$K and $^{53}$Ca are presented, and their implications with respect to the goodness of the \textit{N}=32 sub-shell closure discussed.   
\end{abstract}


\section{Introduction}

Across the nuclear landscape there is evidence that the magic numbers that exist along the line of stability may be eroded with the addition of protons or neutrons.  Approaching the driplines, nucleon single-particle states are reordered, and new shell closures may also develop.  The region surrounding $^{52}$Ca is one where such a reordering has been observed.  The systematic variation of the 2$^{+}$ energies in the neutron-rich isotopes of $_{20}$Ca, $_{22}$Ti and $_{24}$Cr [1-3] shows the anticipated peak at neutron magic number \textit{N}=28, and an additional rise at neutron number \textit{N}=32, suggesting a sub-shell closure. In this region, the energy splitting between the \textit{$\nu$p}$_{3/2}$ and \textit{$\nu$p}$_{1/2}$ spin-orbit partners and the large energy separation of the next available \textit{$\nu$f}$_{5/2}$ state stabilizes the low-energy structure of nuclei.  The stabilization of the \textit{N}=32 sub-shell is most pronounced in the Ca isotopes, and is reduced in nuclei above \textit{Z}=20, due to the strong attractive $\pi$-$\nu$ monopole interaction between the $\pi$\textit{f}$_{7/2}$ and $\nu$\textit{f}$_{5/2}$ single-particle states [4].    

$\beta$ decay and isomeric studies of nuclei surrounding $^{52}$Ca were performed using fast radioactive beams produced by fragmentation at the National Superconducting Cyclotron Laboratory (NSCL).  A primary beam of $^{76}$Ge at an energy of 130 MeV per nucleon was impinged on a thick (352 mg/cm$^{2}$) Be target.  The resulting fragments were transported through the A1900 separator [5], which was operated at its maximum momentum acceptance of 5\%.  Fragments were studied using the NSCL Beta Counting System [6], surrounded by 16 high-purity germanium detectors from the NSCL Segmented Germanium Array (SeGA) [7].  With this arrangement, the simultaneous study of both short-lived isomers, and $\beta$-decay of $\sim$20 nuclei near $^{52}$Ca was possible.  We report here on the low-energy structures of $^{54}_{21}$Sc$_{33}$, $^{50}_{19}$K$_{31}$ and $^{53}_{21}$Sc$_{32}$. 

\section{Results and discussion}

\begin{figure}[!hbtp]
\begin{center}
\includegraphics[width=12cm]{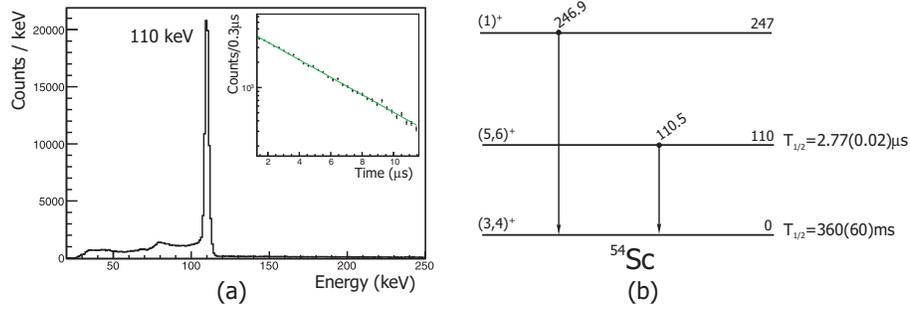}
\caption{(a) Prompt $\gamma$-ray spectrum following implantation of $^{54}$Sc fragments.  Gamma rays were observed for 20 $\mu$s following an implantation event.  The fitted isomer half-life curve yields a half-life value of 2.77(0.02) $\mu$s.  (b) Proposed low-energy level scheme for $^{54}$Sc, including 247 keV state identified in $\beta$ decay [9].}
\label{fig:sc54}
\end{center}
\end{figure}

A 110 keV isomeric transition in $^{54}$Sc was previously observed by Grzywacz \textit{et al.} [8].  The transition was assigned E2 multipolarity based on a comparison of the deduced 7(5) $\mu$s half-life with Weisskopf estimates.  In the present work, over $\sim$650,000 implantations of $^{54}$Sc were observed, and the prompt 110 keV isomeric $\gamma$-ray was confirmed, as depicted in Fig.\ref{fig:sc54}a.  With the high level of statistics available, a decay curve (see Fig.\ref{fig:sc54}a inset) was produced.  The new half-life value of 2.77(0.02) $\mu$s confirms the E2 assignment of this transition. Based on observed $\beta$ feeding from Mantica \textit{et al.} [9] and a (3,4)$^{+}$ ground state, the spin and parity of this isomeric state is limited to (5,6)$^{+}$.  The proposed low-energy level scheme for $^{54}$Sc is presented in Fig.\ref{fig:sc54}b.  The low-energy structure can provide important information on the $\nu$\textit{p}$_{1/2}$ - $\nu$\textit{f}$_{5/2}$ spacing and the possible development of a shell gap at \textit{N}=34.

\begin{figure}[!hbtp]
\begin{center}
\includegraphics[width=12cm]{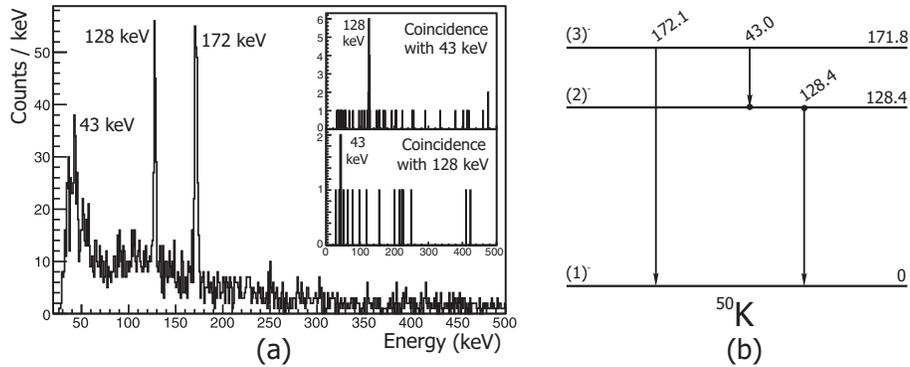}
\caption{(a) Prompt $\gamma$-ray spectrum following implantation of $^{50}$K.  $\gamma$-rays were observed for 20 $\mu$s following an implantation event.  (b) Possible low-energy level scheme for $^{50}$K.  The ordering of the 43-128 keV cascade is not uniquely determined from the present data.
}
\label{fig:k50}
\end{center}
\end{figure}

Isomerism in $^{50}$K was first reported by Lewitowicz \textit{et al.} [10].  In the present results, three $\gamma$-ray transitions were observed in the fragment-$\gamma$ spectrum (see Fig.\ref{fig:k50}a) for $^{50}$K based on $\sim$20,000 implantation events.  The 128 keV and 43 keV transitions were found to be in coincidence.  Additionally, the cross-over transition at 172 keV was observed.  The half-life of this isomer is $<$ 500 ns based on the time spectrum gated on the three isomeric $\gamma$-rays.  Comparison to Weisskopf estimates suggests that such a short-lived 172 keV transition is necessarily E2 in nature.  Considering this, and the observation that the cascade competes strongly with the direct 172 keV transition, serious doubt is cast upon the 0$^{-}$ spin parity assignment for the ground state of this nucleus, as previously suggested from $\beta$ decay [11].  This isomer may be better explained by considering a 1$^{-}$ ground state, arising from a ($\pi$s$_{1/2}$)$^{-1}$($\nu$p$_{3/2}$)$^{-1}$ configuration (see Fig.\ref{fig:k50}b).

\begin{figure}[!hbtp]
\begin{center}
\includegraphics[width=8cm]{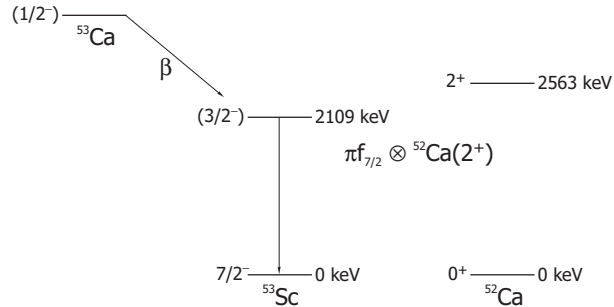}
\caption{Low-energy level scheme for $^{53}$Sc.  Comparison to levels in $^{52}$Ca suggest that $^{53}$Sc can be described by weak coupling of the odd proton to a $^{52}$Ca core.}
\label{fig:sc53}
\end{center}
\end{figure}

Over $\sim$23,000 $\beta$ decay events for $^{53}$Ca permitted identification of a single $\gamma$-ray at 2109 keV that was assigned as a transition in $^{53}$Sc, which has a single proton outside of $^{52}$Ca.  Assuming a robust \textit{N} = 32 subshell closure, the first excited state in $^{53}$Sc should arise from the coupling of the odd \textit{f}$_{7/2}$ proton to the 2$^{+}$ first excited state in $^{52}$Ca.  Such coupling produces a quintet of states, the lowest of which is the 3/2$^{-}$ state.  The 3/2$^{-}$ level should have an energy comparable to the 2$^{+}$ level in $^{52}$Ca (see Fig.\ref{fig:sc53}), and would be the only member of the quintet to be populated directly by $\beta$ decay from the 1/2$^{-}$ $^{53}$Ca ground state.  

\section{Acknowledgements}

This work was supported in part by the NSF under Grant No. PHY06-06007 and by the U.S. DOE, Office of Nuclear Physics, under contracts No. DE-AC02-06CH11357 (ANL) and DEFG02-94ER40834 (U. Maryland) and by the Polish Scientific Committee grant 1PO3B 059 29.  HLC and GFG would like to acknowledge financial support from NSERC of Canada.

\end{document}